\RequirePackage[T1]{fontenc}
\documentclass[conference]{IEEEtran}
\IEEEoverridecommandlockouts %
\usepackage{cite}
\usepackage{amsmath,amssymb,amsfonts}
\usepackage{graphicx}
\usepackage{textcomp}
\usepackage{url}
\usepackage{booktabs}
\usepackage{placeins}

\begin{document}
\title{Non-Commutative State Tracking with Input-Dependent Low-Rank Updates in Mamba-3}
\author{Hiroki Fujii and Masaki Yamakita\thanks{\raggedright The authors are with
Department of Systems and Control Engineering, Institute of Science Tokyo,
Tokyo, Japan. E-mail:\newline
\texttt{fujii.h.8f58@m.isct.ac.jp},\newline
\texttt{yamakita@ac.ctrl.titech.ac.jp}.}\thanks{This work has been submitted to the IEEE for possible publication. Copyright may be transferred without notice, after which this version may no longer be accessible.}}

\maketitle

\begin{abstract}
State tracking from sequential observations can require both retaining information and updating it by composing observed operations.
We extend Mamba-3's diagonal transition with an input-dependent low-rank reflection term to support non-commutative state tracking, in which the order of operations matters.
The rank-one update couples state coordinates along an input-dependent direction, enabling non-diagonal state transitions within a single Mamba-3 block.
The extension retains Mamba-3's trapezoidal input injection, rotary position embeddings (RoPE), and readout.
For training, we adapt chunkwise computation to parallelize the proposed recurrence within each chunk.
Experiments cover group word problems with discrete inputs and a shell game with continuous observations, in which a policy is trained by behavioral cloning.
Among the models selected for their strong performance under fixed timing, the proposed model maintains higher tracking success on longer swap sequences in the shell game with continuous observations and timing jitter.
These experiments show that the proposed method achieves high accuracy on the evaluated non-commutative tracking tasks, improving on standard Mamba-3.
The extension thus offers a Mamba-3-based approach to non-commutative state tracking.

\end{abstract}

\section{Introduction}
\label{sec:intro}
State tracking involves retaining information and, when the underlying state changes, updating it from newly observed operations. For example, an object's initial location no longer identifies its current position after it has been exchanged with another object. State-space models (SSMs) encode histories in time linear in sequence length, and Mamba-3~\cite{mamba3} improves state tracking through complex-valued state updates. SSM-based history encoders have also been applied to imitation learning under partial observation, where past observations and operations are needed to select an action. Evaluating such models therefore requires distinguishing information retention from operation composition.

Tracking exchanged objects requires ordered composition of permutations, motivating non-commutative state transitions. Asymptotically, word problems for fixed finite non-solvable groups are $\mathrm{NC}^1$-complete~\cite{barrington}, whereas the fixed-depth Transformers and diagonal-transition SSMs analyzed by Merrill et al.~\cite{merrill} lie in uniform $\mathrm{TC}^0$ under logarithmic precision. If $\mathrm{TC}^0\ne\mathrm{NC}^1$, as conjectured, the analyzed architectures cannot solve all such words exactly at arbitrary length. Empirically, Merrill et al.~\cite{merrill} found that Mamba required increasing depth to achieve high accuracy on longer permutation sequences. With layer width held fixed, using more layers for these non-commutative tasks also increases the parameter count.

Input-dependent non-diagonal transitions are an established approach to state tracking. We incorporate this approach into Mamba-3 by adding an input-dependent reflection term $N(x_t)$ to its diagonal scan transition. The resulting transition has an input-dependent normal-plus-low-rank (NPLR) structure. For brevity, we refer to the model as Mamba-3 + NPLR. To evaluate its state-tracking capabilities, we test discrete operation composition using group word problems over the cyclic group $\mathbb{Z}_5$ and the symmetric group $S_5$. Furthermore, a shell game tests whether models can learn to compose observed swaps from continuous observations and demonstrated actions.

Our contributions and supporting evidence are as follows.
(1) We propose an input-dependent reflection update that enables non-commutative state tracking within a single Mamba-3 block. We demonstrate its effectiveness on the $S_5$ generator task and a continuous-observation shell game, where standard Mamba-3 retains tracking errors under the main training conditions.
(2) We provide a Mamba-3-based option for non-commutative state tracking by combining its input and readout mechanisms with reflection-based state updates. This expands the choice of architectures for these tasks beyond existing delta-based models, whose relative performance depends on the task.
(3) We demonstrate an advantage over Gated DeltaNet (GDN), DeltaProduct, and its gated variant Gated DeltaProduct (GDP) in state tracking from continuous observations under timing jitter, suggesting potential applications to dynamic physical environments.

Section~\ref{sec:related} reviews related work, and Section~\ref{sec:rx} presents the proposed method.
Section~\ref{sec:bench} defines the evaluation tasks, and Section~\ref{sec:result} reports the experimental results.
Section~\ref{sec:discussion} discusses the findings and limitations, and Section~\ref{sec:conclusion} concludes the paper.
The appendixes provide the chunkwise derivation, experimental settings, and additional results.

\section{Related Work}
\label{sec:related}

\paragraph{Computational theory of state tracking}
For fixed finite solvable groups, operation sequences can be evaluated by circuits with modular counting gates whose depth does not grow with sequence length~\cite{barringtontherien}. Related work constructs shallow Transformers for solvable automata~\cite{liu}.
Empirical studies of permutation composition examine whether language models learn shortcuts or track the underlying state~\cite{listate}.
The difference in expressivity between diagonal and non-diagonal input-dependent transitions has been characterized theoretically~\cite{cirone}.

\paragraph{Input-dependent transitions and Mamba extensions}
Building on HiPPO~\cite{hippo}, S4~\cite{s4} uses a diagonal-plus-low-rank structure.
Subsequent work explores diagonal parameterizations~\cite{dss,s4d} and the input-dependent updates of Mamba-1/2~\cite{mamba,mamba2}.
Within this line of work, Grazzi et al.~\cite{grazzi} showed that allowing negative eigenvalues in the state-transition matrices of Mamba and DeltaNet enables these models to learn parity, a commutative state-tracking task based on repeated sign changes.
Yang et al.~\cite{deltanet} use a memory-efficient representation of Householder products to accelerate DeltaNet training.
DeltaProduct~\cite{deltaproduct} composes multiple generalized Householder reflections per step and solves $S_5$ word problems.
RWKV-7~\cite{rwkv7} also uses input-dependent diagonal-plus-low-rank state transitions, with vector-valued gating and update rates.
GDN~\cite{gdn} combines Mamba-2's scalar gate and a rank-1 delta rule multiplicatively.
Our earlier work~\cite{bim}, motivated by Koopman bilinear systems, introduced a multiplicative interaction between the input and state through gating but exhibited stability issues.

Bouhadjar et al.~\cite{selectivbench} compare Mamba and Mamba-2 with GDN and GDP and report task-dependent relative performance.
On their benchmarks, Mamba-based models perform well on memory and context-dependent selection, whereas GDN and GDP perform well on noise rejection and generalization across long noise intervals.
Their study does not include Mamba-3, but highlights the need to assess practical performance beyond transition expressivity.
Motivated by this task dependence, we provide a Mamba-3-based implementation for non-commutative state tracking.
The Mamba-3 study~\cite{mamba3} evaluates parity and modular arithmetic but does not evaluate word problems over non-solvable groups.
We address this setting by evaluating the low-rank extension on $S_5$ word problems.

\paragraph{Application to imitation learning with history}
Action Chunking with Transformers (ACT)~\cite{act} uses action-chunk prediction for real-robot imitation.
The robomimic framework~\cite{robomimic} provides recurrent baselines, including behavioral cloning with a recurrent neural network (BC-RNN) with a two-layer LSTM.
MaIL~\cite{mail} builds imitation-learning policies with Mamba.
MTIL~\cite{mtil} sequentially encodes full trajectories with Mamba-2~\cite{mamba2} and reports substantial improvements over ACT on multistage tasks.
RoboSSM~\cite{robossm} uses SSMs for in-context imitation.
These applications motivate studying whether a Mamba backbone can track an object's current location as successive swaps change it. We examine operation composition through group word problems and a shell game with continuous observations and demonstrated actions as a step toward such applications.

\section{Proposed Method}
\label{sec:rx}

\subsection{Input-Dependent Low-Rank State Updates}
\label{sec:nplr}
The proposed Mamba-3 extension with an input-dependent NPLR transition (Mamba-3 + NPLR) adds a low-rank reflection term to the diagonal scan transition. The block's rotary position embeddings, trapezoidal correction, normalization, and gates are retained.

Let $x_t\in\mathbb R^{d_{\rm model}}$ denote the layer input, obtained by embedding the observation. Following Mamba-3, we use $H$ heads, with $P$ channels per head and a state dimension of $n$ per channel. Each channel has its own state $h_t\in\mathbb R^n$, giving $Pn$ state elements per head. The matrix $I\in\mathbb R^{n\times n}$ is the identity. Unless otherwise stated, equations describe a single head and channel, with their indices omitted.

We add the reflection term to the discretized state transition while leaving Mamba-3's input injection terms unchanged. The state update is
\begin{equation}
\begin{aligned}
  h_t ={}& \big(d_t I + N(x_t)\big)h_{t-1}
       + \lambda_t\Delta_t\widetilde B_t u_t\\
       &{}+(1-\lambda_t)\Delta_t d_t\widetilde B_{t-1}u_{t-1}.
\end{aligned}
  \label{eq:frame}
\end{equation}
The scalars $d_t,\Delta_t,\lambda_t$ follow Mamba-3 and are shared within each head, with
\begin{equation}
\label{eq:decay-parameterization}
\begin{aligned}
  a_t &= \max\{\mathrm{softplus}(W_a x_t),a_{\min}\},\\
  \Delta_t &= \mathrm{softplus}(W_\Delta x_t+b_\Delta),
  \qquad d_t=e^{-a_t\Delta_t}
\end{aligned}
\end{equation}
For a fixed head, $W_a,W_\Delta,W_\lambda\in\mathbb R^{1\times d_{\rm model}}$ are learned row projections. The scalar $b_\Delta$ is a learned bias, and $a_{\min}$ is a lower bound on the decay rate.
Here, $u_t$ is the channel-specific scalar value projection of $x_t$, and $\widetilde B_t\in\mathbb R^n$ is the injection vector defined below. The input-dependent coefficient $\lambda_t=\sigma(W_\lambda x_t)$ controls the trapezoidal correction, where $\sigma$ is the logistic sigmoid. The contribution from the preceding input is set to zero only at the start of the sequence.

Following Mamba-3, let $\widehat B_t,\widehat C_t\in\mathbb R^n$ be the projected and normalized $B/C$ vectors after adding head-specific biases. The vectors before bias addition are shared across heads, and the resulting vectors are shared by the $P$ channels within each head. Cumulative RoPE and the readout are
\begin{equation}
\label{eq:rope-readout}
\begin{aligned}
 \Theta_t&=\sum_{j=1}^{t}\Delta_j\omega_j,\qquad Q_t=\mathcal R(\Theta_t),\\
 \widetilde B_t&=Q_t\widehat B_t,\qquad \widetilde C_t=Q_t\widehat C_t,\\
 y_t&=\widetilde C_t^\top h_t+D u_t
\end{aligned}
\end{equation}
For $m\le n/2$ rotated coordinate pairs, $\omega_t,\Theta_t\in\mathbb R^m$ and $Q_t\in\mathbb R^{n\times n}$. The angular velocity follows Mamba-3 as $\omega_t=\pi\tanh(W_\omega x_t)$, where $W_\omega\in\mathbb R^{m\times d_{\rm model}}$ is learned and shared across heads. The operator $\mathcal R$ rotates adjacent coordinate pairs and acts as the identity on the remaining coordinates. The implementation applies $Q_t$ without materializing a dense matrix. The learned scalar $D$, also shared within the head, controls the skip connection, and $y_t\in\mathbb R$ is the channel-specific readout before output normalization, gating, and projection.

Additional linear projections of the same input generate the reflection coefficient and direction. We use a rank-1 term corresponding to a generalized Householder update,
\begin{equation}
  N(x_t)=-\beta_t k_tk_t^\top,\qquad R=1,
  \label{eq:reflection}
\end{equation}
where $k_t=Q_t\widehat k_t\in\mathbb R^n$, $\widehat k_t=W_kx_t/\max\{\|W_kx_t\|,\varepsilon\}$, and $\beta_t=2\sigma(w_\beta^\top x_t+b_\beta)\in(0,2)$. For a fixed head, $W_k\in\mathbb R^{n\times d_{\rm model}}$ and $w_\beta\in\mathbb R^{d_{\rm model}}$ are learned projections. The scalar $b_\beta$ is a learned bias, and $\varepsilon=10^{-12}$ stabilizes the normalization. Both $k_t$ and $\beta_t$ are shared by the $P$ channels within each head. This term modifies the previous state along an input-dependent direction, allowing attenuation or sign reversal.

The direction $k_t$ uses the same cumulative rotation as $B/C$. The dependence on the accumulated phase is implicit in the notation $N(x_t)$. The no-RoPE ablation sets $Q_t=I$ for $B/C$ and $k_t$ while retaining the reflection term.

Removing the low-rank term ($R=0$) recovers the Mamba-3 state update. Our main comparisons use standard Mamba-3 with the same task-specific input and output mappings, training conditions, and evaluation protocol as the proposed model.

\subsection{Non-Commutative Transitions from a Reflection Term}
\label{sec:noncommutative}
For each head, write the transition in scan coordinates as $A_t=d_tI+N_t$, where $N_t=N(x_t)$ and $d_t$ is a scalar. Defining the commutator as $[A_i,A_j]=A_iA_j-A_jA_i$ gives
\begin{equation}
\label{eq:commutator-general}
  [A_i,A_j]=[N_i,N_j]
\end{equation}
Without a low-rank term, $N_t=0$ and the transitions commute. With the reflection term $N_t=-\beta_t k_tk_t^\top$,
\begin{equation}
\label{eq:commutator-reflection}
  [A_i,A_j]=\beta_i\beta_j(k_i^\top k_j)
  \bigl(k_i k_j^\top-k_j k_i^\top\bigr).
\end{equation}
Thus, when $\beta_i\beta_j\ne0$ and $k_i,k_j$ are neither parallel nor orthogonal, the transitions do not commute and can represent order-dependent state updates. This property concerns only the transition, excluding input injection and readout. It does not guarantee successful learning of a group word problem.

\subsection{Chunkwise Computation for Training}
\label{sec:chunk}
The reflection term can be applied to an $n$-vector $h$ as $-\beta_t k_t(k_t^\top h)$, requiring $O(n)$ operations per channel and time step in a sequential update. This cost excludes coefficient generation, input injection, and readout. For training, we adapt the chunkwise computation used for delta-rule models~\cite{deltanet}. Yang's technical note~\cite{yangdeltanetblog} provides an accessible account of the underlying algorithm.

We apply a lower-triangular solve to the scalar decay, rank-1 reflection, and trapezoidal injection in \eqref{eq:frame}. Within each chunk, the $P$ channels of a head share one coefficient matrix and use channel-specific right-hand sides. The final state is passed to the next chunk, and the preceding injection factors are retained across chunk boundaries. This computation exactly rearranges the recurrence. Appendix~\ref{app:chunk_derivation} gives the derivation and the direct readout that avoids reconstructing all intermediate states.

After coefficient generation, the forward computation of state updates and readouts requires $O(LnP+Lc(n+P))$ operations per sequence and head, where $L$ is the sequence length and $c$ is the chunk length. For a fixed chunk length $c$, this cost is linear in $L$.

\section{Evaluation Tasks}
\label{sec:bench}

\subsection{Group Word Problems}
\label{sec:word}
Following prior work~\cite{deltaproduct}, we use the term group word problem for the task of predicting the group element represented by each input prefix. Given a sequence $g_1 \dots g_L$ over a group $G$, the model classifies the prefix product $s_t = g_t \circ s_{t-1}$ at each position, starting from the identity $s_0=e$. The target is the product itself, rather than a binary decision about whether it equals the identity. Correctly predicting the product also allows the identity test to be answered.

We use the cyclic group $\mathbb{Z}_5$ of order five for commutative tracking and the symmetric group $S_5$ of permutations of five elements for non-commutative composition. The task over $\mathbb{Z}_5$ is cyclic addition modulo five, $s_t = (s_{t-1} + g_t) \bmod 5$, with $s_0 = 0$. Here $g_t\in\mathbb Z_5$ is the input increment, and the result depends only on the sum of the increments. For $S_5$, the inputs are adjacent transpositions, and the prefix product is classified into one of 120 classes. We refer to this setting as $S_5$ (generators).

Training sequences contain at most 64 group elements. We evaluate accuracy at each sequence position and at the final position, using sequences of length 64 and longer. Training procedures and chance levels are given in Appendix~\ref{app:task_settings}.

\subsection{Shell Game}
\label{sec:shell}
\begin{figure}[t]
  \centering
  \includegraphics[width=\linewidth]{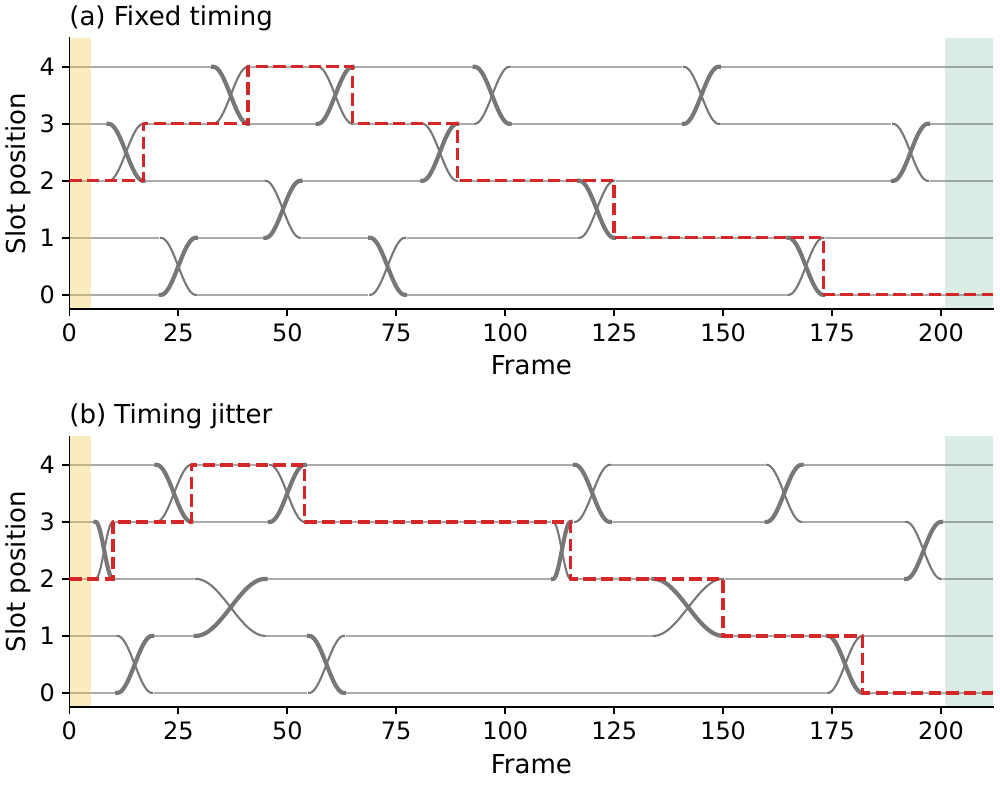}
  \caption{Shell-game illustration with the same 12 swaps under (a) fixed timing and (b) timing jitter. Cups remain stationary in windows without a swap. Gray curves show cup positions. Thick segments indicate the lifted cup. The red dashed line shows the ball-containing slot, updated at swap completion. Yellow and green intervals mark the reveal and reach phases. Timing durations are chosen for illustration, with both episodes spanning 213 frames.}
  \label{fig:shell}
\end{figure}

The ball's location is revealed only during an initial reveal phase. Five visually identical cups then undergo $K$ adjacent swaps, with $K$ sampled uniformly from $\{0,\dots,16\}$. One cup is lifted during each swap, making the exchange observable. The policy receives a 16-dimensional observation consisting of cup coordinates sorted by position, a one-hot reveal indicator, and a cue. Cup identities are not observed. After the cue, the policy outputs a reach position toward the cup containing the ball (Fig.~\ref{fig:shell}).

Because the swaps are visible but the ball is not revealed again, correct tracking requires composing the transpositions in the swap history. This task tests whether the composition required by group word problems can be learned from continuous observations and continuous action targets.

We evaluate the shell game in two stages. First, we compare models with fixed waiting and motion durations. Second, we restrict the evaluation to models that perform well in the first stage and introduce random jitter in swap intervals and speeds by varying these durations within each episode.

In the fixed-timing setting, models are trained by behavioral cloning for 20{,}000 updates with $K \le 16$ and sequence length $T{=}213$. We call evaluation within this training range the standard condition. Direct training uses the full swap-count range from the outset. The loss is the squared reach error over all time steps, with greater weight on the steps after the cue.

A prediction is successful if the slot nearest to the model's output during the response interval is the correct slot. For the fixed-timing comparison, success rates are reported separately for each $K$ using 4,096 episodes per run in each evaluation condition. In the extended condition, trained models are evaluated at $K \le 32$ and $T{=}405$. This jointly tests extrapolation in the number of swaps $K$ and sequence length $T$. Each fixed-timing episode has $W$ windows of four quiet frames and eight motion frames, plus five initial reveal frames, four final quiet frames, and 12 response frames. This gives 213 frames for 16 windows and 405 frames for 32 windows. Only $K$ windows contain swaps. Inactive windows retain their full duration.

In addition to the 20\% chance level of a constant prediction, we use a diffusion-based no-tracking prior that depends only on the revealed location and $K$. Success at $K=0$ indicates retention, whereas evidence of composition requires exceeding this prior at $K>0$. Observations do not depend on the policy's outputs, so this evaluation does not measure closed-loop covariate shift. Input dimensions, the sequence structure, and loss weights are provided in Appendix~\ref{app:task_settings}.

In the timing-jitter setting in Fig.~\ref{fig:shell}(b), the models specified in Section~\ref{sec:model_comparison} are trained from scratch with up to 16 swaps. We evaluate longer swap sequences under the training-time timing distribution using the same 1,024 test episodes per condition across all compared models and all three runs. The timing distributions are specified in Appendix~\ref{app:task_settings}.

\subsection{Models Compared}
\label{sec:model_comparison}
Across the group tasks and the fixed-timing shell game, we compare Mamba-3 + NPLR with standard Mamba-3, DeltaProduct, the GDN extension, and Input-Dependent S4 (IDS4)~\cite{merrill}. For the timing-jitter shell game, we compare Mamba-3 + NPLR, DeltaProduct ($n_h=4$), and GDN, each of which achieves high success in at least two fixed-timing runs. We also include the no-RoPE ablation of Mamba-3 + NPLR in this setting. As a supplemental control, we evaluate GDP across the group and shell-game tasks, using $n_h=4$ for timing jitter (Appendixes~\ref{app:dp_settings} and~\ref{app:dp_gdp}).

Mamba-3 + NPLR uses a single block. We evaluate standard Mamba-3 with one, two, or four blocks, and DeltaProduct, the GDN extension, and IDS4 with one layer each. DeltaProduct uses two or four reflection updates per input. Parameter counts are not matched across architectures. Mamba-3 + NPLR uses fewer parameters than DeltaProduct, GDP, GDN, and IDS4 on both tasks. Table~\ref{tab:sym} reports the shell-game model sizes, and model-specific configurations are given in Appendix~\ref{app:model_settings}.

For each architecture and block count, the core block configuration and recurrent-state size are held fixed across the two tasks. Only the task-specific input and output mappings differ. Group tasks use learned token embeddings and classification heads, whereas the shell game uses a linear projection of the 16-dimensional observations and an action readout. Each model is trained separately for each task using the corresponding objective and training procedure.

\section{Experimental Results}
\label{sec:result}

\subsection{Learning Group Word Problems with Low-Rank Updates}
\label{sec:nplrres}
Mamba-3 + NPLR achieves perfect final-position accuracy at the training length on $S_5$ (generators) for all three runs, whereas one-block Mamba-3 without the reflection term remains at chance level (Table~\ref{tab:nplr}). Increasing the depth of Mamba-3 improves accuracy, but even four blocks do not achieve perfect tracking at the training length (Table~\ref{tab:nplr}, three runs). GDN~\cite{gdn} with its coefficient range extended to $\beta \in (0,2)$ and DeltaProduct~\cite{deltaproduct} also achieve high accuracy on the same task, as does IDS4.

On $\mathbb{Z}_5$, the proposed model and IDS4 achieve high accuracy at the training length. Both DeltaProduct configurations exceed chance level but, like the GDN extension, still make errors at the training length (Table~\ref{tab:nplr}). These results apply to the evaluated configurations and training conditions.

\begin{table}[t]
  \centering
  \caption{Final-position accuracy (\%) on group word problems (arithmetic mean over three runs). Models are trained at length 64. For DeltaProduct, $n_h$ is the number of reflection updates per input. Values are rounded to two decimal places, so 100.00\% does not necessarily indicate perfect accuracy. Bold values satisfy $\ge 95\%$ before rounding.}
  \label{tab:nplr}
  \footnotesize\setlength{\tabcolsep}{2pt}
  \begin{tabular}{lccc}
    \toprule
    Model & \shortstack{$\mathbb{Z}_5$\\$L=64$} & \shortstack{$S_5$\\$L=64$} & \shortstack{$S_5$\\$L=128$} \\
    \midrule
    Mamba-3 (1 block) & \textbf{99.93} & 0.78 & 0.91 \\ %
    Mamba-3 (2 blocks) & \textbf{100.00} & 5.53 & 1.11 \\ %
    Mamba-3 (4 blocks) & \textbf{100.00} & 52.86 & 2.28 \\ %
    \textbf{Mamba-3 + NPLR (Proposed)} & \textbf{100.00} & \textbf{100.00} & \textbf{99.61} \\ %
    DeltaProduct ($n_h{=}2$) & 89.78 & \textbf{100.00} & \textbf{99.28} \\ %
    DeltaProduct ($n_h{=}4$) & \textbf{99.48} & \textbf{100.00} & \textbf{99.87} \\ %
    Gated DeltaNet ($\beta\in(0,2)$) & 84.44 & \textbf{100.00} & \textbf{100.00} \\ %
    IDS4 & \textbf{100.00} & \textbf{100.00} & \textbf{100.00} \\ %
    \bottomrule
  \end{tabular}

\end{table}

\begin{figure}[!t]
  \centering
  \includegraphics[width=\linewidth]{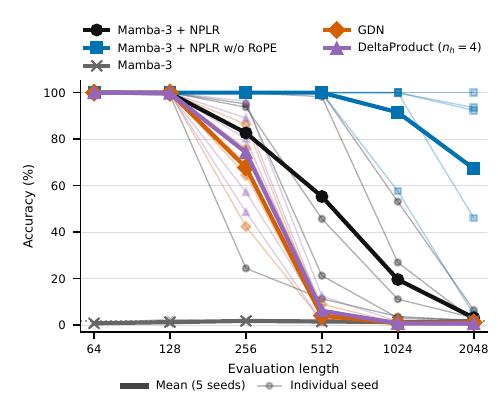}
  \caption{Length extrapolation on $S_5$ (generators), measured by final-position accuracy (\%). The training length is 64. Thick lines show five-run means. Thin lines of the same color show individual runs. The dotted line is the uniform-prediction baseline over the 60 candidates allowed by parity ($1/60$).}
  \label{fig:extrap}
\end{figure}

For the $S_5$ extrapolation comparison in Fig.~\ref{fig:extrap}, all models are trained for 60,000 updates with a final training length of 64 and evaluated at lengths from 64 to 2048. At each length, all models and runs use the same 512 test sequences, generated independently of training. Mamba-3 has one block, and DeltaProduct uses $n_h=4$.

The proposed model maintains high accuracy at twice the training length but degrades on longer words. Other models also exhibit extrapolation limits and variation across runs, so success at the training length alone does not determine performance on longer sequences (Fig.~\ref{fig:extrap}, Table~\ref{tab:extrapdetail}). Disabling RoPE in the proposed model yields perfect accuracy through evaluation length 512 for all five runs (Table~\ref{tab:extrapdetail}(c)). At greater lengths, accuracy declines and differences across runs remain.

\subsection{State Tracking in the Shell Game}
\label{sec:exist}
\label{sec:hierres}
\begin{figure*}[t]
  \centering
  \includegraphics[width=\textwidth]{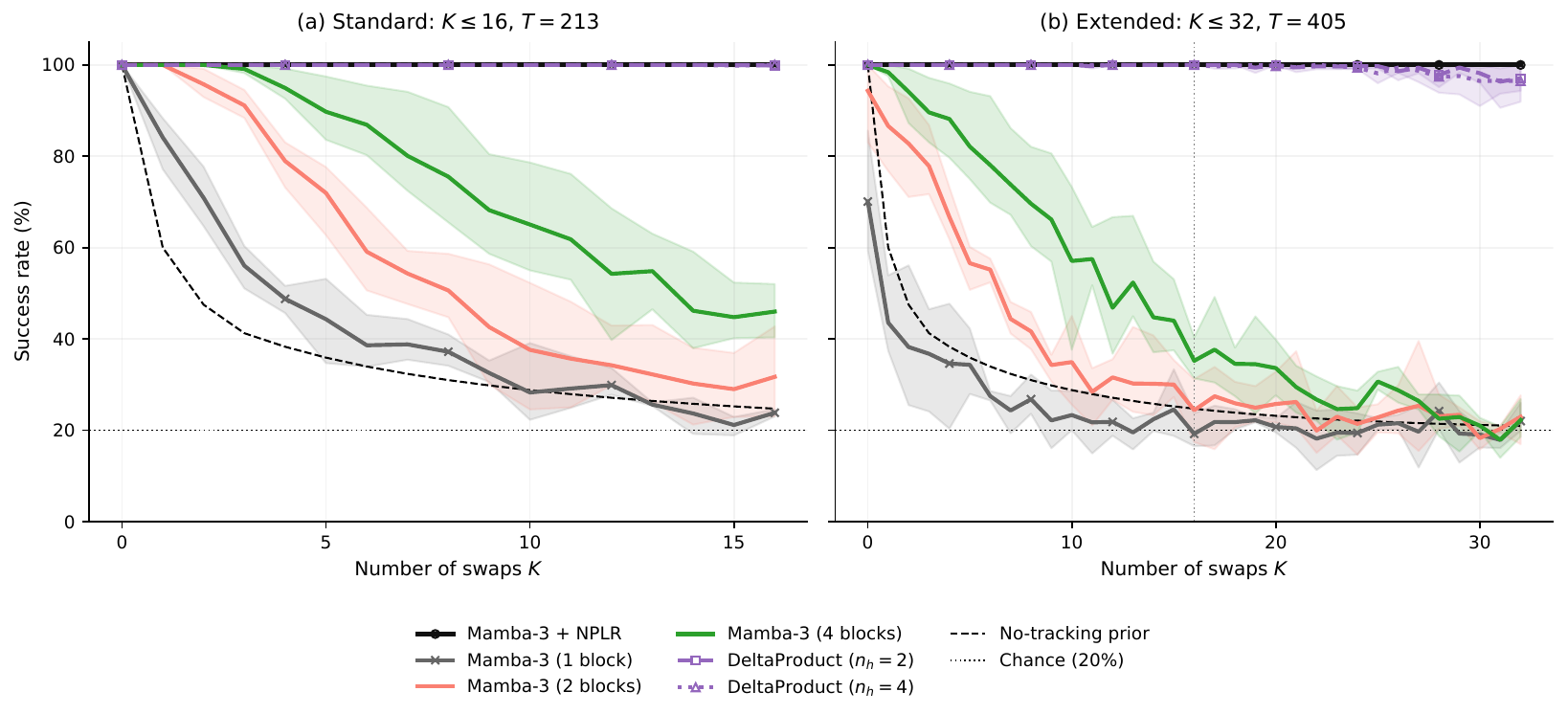}
  \caption{Shell-game success rates by number of swaps in the standard (left, $T=213$, $K\leq16$) and extended (right, $T=405$, $K\leq32$) conditions. Lines show means over three runs, and bands span the minimum and maximum. All models use direct training for 20{,}000 updates. The same checkpoints are evaluated on 4{,}096 episodes in each condition. The black dashed line is the no-tracking baseline, the dotted line is the 20\% chance level, and the vertical line in the right panel marks the training limit $K=16$.}
  \label{fig:rx}
\end{figure*}

After direct training, the proposed model achieves perfect success in both the standard and extended fixed-timing conditions for all three runs (Fig.~\ref{fig:rx}). DeltaProduct also achieves high success rates in both conditions, with a small number of errors. Comparison with a standard one-block Mamba-3 using the same task-specific input and output mappings assesses the contribution of the low-rank term. Success rates at intervals of four swaps, including GDN, IDS4, and GDP results not shown in Fig.~\ref{fig:rx}, are reported in Tables~\ref{tab:shell_full_standard} and~\ref{tab:shell_full_extended} in Appendix~\ref{app:additional_results}.

Under the swap-count curriculum, GDN also achieves near-perfect success in the standard condition. Table~\ref{tab:selected} in Appendix~\ref{app:additional_results} reports the corresponding comparison across models.

\subsection{Generalization under Timing Jitter}
\label{sec:jitter}
In applications such as imitation learning, action durations and waiting times between actions can vary. Following the protocol in Section~\ref{sec:shell}, we evaluate extrapolation up to 128 swaps.

\begin{figure}[t]
  \centering
  \includegraphics[width=\linewidth]{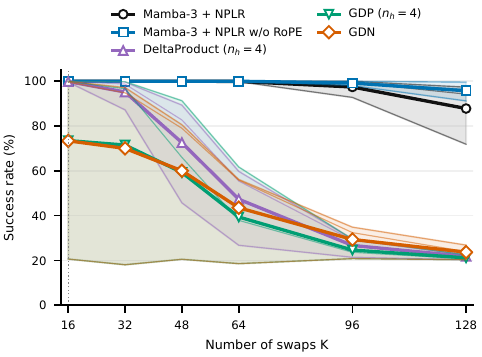}
  \caption{Shell-game success in the timing-jitter setting after training with up to 16 swaps. Thick lines show means over all three runs, thin lines show individual runs, and shading spans the minimum and maximum. One run each for GDN and GDP remains near chance across the evaluated swap counts. Models use the same 1,024 test episodes at each swap count. The no-RoPE ablation sets $Q_t=I$ for $B/C$ and $k_t$. The vertical line marks the training limit. Motion and waiting-duration distributions are unchanged across all six evaluation points.}
  \label{fig:shell_jitter}
\end{figure}

Mamba-3 + NPLR achieves perfect success through 48 swaps in all three runs. Its mean success remains above 95\% through 96 swaps and decreases to 87.83\% at 128 swaps, compared with 21.81\% for DeltaProduct and 21.16\% for GDP (Fig.~\ref{fig:shell_jitter}). One GDP run remains near chance throughout evaluation, and the other two also decline toward chance on longer sequences. For GDN, two runs achieve over 99\% success at $K=16$, but both deteriorate substantially on longer sequences. The three-run mean for GDN decreases to 23.70\% at $K=128$. Performance also varies across runs at longer lengths. These results show stronger, but finite, extrapolation in the evaluated timing-jitter setting.

Disabling RoPE yields perfect success through 64 swaps in all three runs and a mean of 95.74\% at 128 swaps. Thus, RoPE does not improve performance in this evaluated setting.

\section{Discussion}
\label{sec:discussion}

\subsection{State Tracking Performance}
\label{sec:discussion_tracking}
Compared with standard Mamba-3 using the same task-specific input and output mappings, direct-training procedure, and evaluation conditions, adding the input-dependent low-rank term improves success rates in the evaluated shell-game conditions (Fig.~\ref{fig:rx}).

The timing-jitter results show differences in extrapolation beyond the training swap count (Fig.~\ref{fig:shell_jitter}). As individual motion and waiting durations follow the same distribution at training and evaluation, the separation emerges when more operations must be tracked over longer sequences. A possible explanation for our timing-jitter results is that the proposed model preserves the tracked state more reliably across successive motion and waiting intervals, reducing the accumulation of errors. GDN and GDP also combine input-dependent decay with reflection-based updates, so these shared mechanisms alone do not explain the difference. The proposed extension additionally retains Mamba-3's trapezoidal input injection and complex-valued state dynamics, providing additional mechanisms for processing temporal variation~\cite{mamba3}. These mechanisms may complement reflection-based tracking, although their benefits depend on the task, as suggested by our RoPE ablation.

The proposed model achieves these results with fewer parameters than DeltaProduct, GDP, GDN, and IDS4 and no larger recurrent state (Table~\ref{tab:sym}). These results are achieved with a single rank-1 reflection update per input, compared with two or four in the evaluated DeltaProduct configurations. This highlights the effectiveness of combining a lightweight reflection update with Mamba-3's existing mechanisms.

\subsection{Architectural Components and Task Performance}
\label{sec:discussion_components}

In Table~\ref{tab:capability}, selectivity denotes input dependence of coefficients, writing, and readout. For IDS4, it refers to the transition matrix. Decay denotes an independent scalar decay gate, reflection a generalized Householder update, and RoPE rotation of $B/C$. The proposed model also applies this rotation to its reflection direction. Models have one layer or block unless noted. Two- and four-block Mamba-3 provide depth comparisons.

$\mathbb{Z}_5$ and $S_5$ (generators) report final-position accuracy at the training length of 64. Shell reports overall success rate in the standard condition ($T=213$, $K\le16$) after 20,000 direct-training updates. Tasks and evaluation are defined in Section~\ref{sec:bench}. Additional evaluation results are given in Appendix~\ref{app:additional_results}.

\begin{table*}[t]
  \centering
  \caption{Component paths and task performance. $\checkmark$/-- denotes the presence or absence of an explicit path, not a verdict on capability or a necessary or sufficient condition for it. Results are three-run means expressed as percentages. Bold values satisfy $\ge 95\%$ before rounding. $^{\ddagger}$The shell-game means for GDN and GDP each combine two perfect runs and one run that predicts only the central slot.}
  \label{tab:capability}
  \footnotesize\setlength{\tabcolsep}{3pt}
  \begin{tabular}{lccccc|ccc}
    \toprule
    & Selectivity & Decay & Reflection & RoPE & Non-diagonal
      & $\mathbb{Z}_5$ & $S_5$ & Shell game \\
    Model & & gate & & ($B/C$) & transition & $L=64$ & $L=64$ & \\
    \midrule
    Mamba-3 (1 block) & $\checkmark$ & $\checkmark$ & -- & $\checkmark$ & -- & \textbf{99.93} & 0.78 & 43.33 \\ %
    Mamba-3 (2 blocks) & $\checkmark$ & $\checkmark$ & -- & $\checkmark$ & -- & \textbf{100.00} & 5.53 & 57.55 \\ %
    Mamba-3 (4 blocks) & $\checkmark$ & $\checkmark$ & -- & $\checkmark$ & -- & \textbf{100.00} & 52.86 & 74.65 \\ %
    \textbf{Mamba-3 + NPLR (Proposed)} & $\checkmark$ & $\checkmark$ & $\checkmark$ & $\checkmark$ & $\checkmark$
      & \textbf{100.00} & \textbf{100.00} & \textbf{100.00} \\ %
    DeltaProduct ($n_h{=}2$) & $\checkmark$ & -- & $\checkmark$ & -- & $\checkmark$
      & 89.78 & \textbf{100.00} & \textbf{99.99} \\ %
    DeltaProduct ($n_h{=}4$) & $\checkmark$ & -- & $\checkmark$ & -- & $\checkmark$
      & \textbf{99.48} & \textbf{100.00} & \textbf{99.99} \\ %
    GDP ($n_h{=}4$) & $\checkmark$ & $\checkmark$ & $\checkmark$ & -- & $\checkmark$ & \textbf{97.46} & \textbf{100.00} & 73.62$^{\ddagger}$ \\ %
    Gated DeltaNet ($\beta\in(0,2)$) & $\checkmark$ & $\checkmark$ & $\checkmark$ & -- & $\checkmark$
      & 84.44 & \textbf{100.00} & 73.62$^{\ddagger}$ \\ %
    IDS4 & $\checkmark$ & -- & -- & -- & $\checkmark$
      & \textbf{100.00} & \textbf{100.00} & 28.03 \\ %
    \bottomrule
  \end{tabular}

\end{table*}

Mamba-3 and GDN both include selectivity and a decay gate, yet their accuracies on $\mathbb{Z}_5$ differ under the evaluated conditions (Table~\ref{tab:capability}). DeltaProduct, GDP, and the GDN extension achieve high accuracy on the generator task, whereas GDP and the GDN extension exhibit substantial variation across runs in the standard fixed-timing shell game (success rates from 20.87\% to 100\%). These comparisons show that component presence alone does not explain task performance.

In the group-composition task, each generator represents the same operation regardless of its position, and the target depends on the ordered composition rather than the timing of the inputs. RoPE may therefore introduce phase dependence that is unnecessary for this task, potentially making extrapolation beyond the training length more difficult (Fig.~\ref{fig:extrap}). Likewise, the correct final slot in the shell game depends on swap composition rather than on motion or waiting durations. The no-RoPE variant matches performance at shorter lengths and achieves higher mean success at longer lengths in the timing-jitter setting (Section~\ref{sec:jitter}). These results provide no evidence of a RoPE benefit in this setting.

\subsection{Applications and Limitations}
\label{sec:discussion_limits}
Tasks such as tracking identical parts under occlusion are potential applications of this construction. They require updating the current correspondence from observed operations and selecting actions based on that correspondence.

The main results for the proposed method are evaluated over three runs. The evaluations are limited to tasks with discrete inputs and low-dimensional observations. Effectiveness with image observations or physical systems has not been tested. In the shell game, model outputs do not affect subsequent observations, so conditions in which actions and observations influence each other are not evaluated.

These evaluations leave open when RoPE's additional temporal flexibility benefits non-commutative state tracking. Future work will examine more realistic imitation-learning tasks that require both composing observed operations and using their timing to select actions, to assess the complementary roles of reflection-based updates and RoPE.

\section{Conclusion}
\label{sec:conclusion}
We presented Mamba-3 + NPLR, which adds an input-dependent reflection term to enable non-commutative state tracking within a single Mamba-3 block. Experiments on $S_5$ (generators) and a continuous-observation shell game demonstrate improved tracking over standard Mamba-3 under the evaluated conditions. By combining Mamba-3's input and readout mechanisms with reflection-based state updates, the proposed construction provides an additional architectural option beyond existing delta-based models. Its advantage over DeltaProduct and GDP, both with $n_h=4$, and GDN with $\beta\in(0,2)$ on longer timing-jitter shell-game sequences illustrates the value of this option and suggests potential for state tracking in dynamic physical environments.

\appendices
\section{Chunkwise State-Update Derivation}
\label{app:chunk_derivation}

We partition the sequence into fixed-length chunks and express the updates within each chunk as a linear system with a lower-triangular coefficient matrix. The final state of each chunk becomes the initial state of the next. The trapezoidal injection uses the preceding factors $\widetilde B_{t-1},u_{t-1}$ even at chunk boundaries. Their contribution is zero only at the start of the full sequence.

Combining the two injection terms in \eqref{eq:frame} into $b_t\in\mathbb R^n$ gives the rank-1 recurrence $h_t=d_th_{t-1}-\beta_t k_t(k_t^\top h_{t-1})+b_t$. Within a chunk, index time by $t=1,\ldots,c$ and denote the incoming state by $h_0$. Define $E_{t,j}=\prod_{\ell=j+1}^{t}d_\ell$, with the empty product equal to one. Let $f_t\in\mathbb R^n$ be the response without reflections, and define scalars $q_t,v_t$ by
\begin{equation}
\label{eq:chunk-definitions}
\begin{aligned}
 f_t&=E_{t,0}h_0+\sum_{j=1}^{t}E_{t,j}b_j,\qquad f_0=h_0,\\
 q_t&=k_t^\top h_{t-1},\qquad v_t=k_t^\top f_{t-1},\\
 M_{tj}&=\begin{cases}
 \beta_j E_{t-1,j}k_t^\top k_j,&j<t,\\
 0,&j\ge t.
 \end{cases}
\end{aligned}
\end{equation}
With $M\in\mathbb R^{c\times c}$, $q=(q_1,\ldots,q_c)^\top\in\mathbb R^c$, and $v=(v_1,\ldots,v_c)^\top\in\mathbb R^c$, the vector $q$ satisfies the unit lower-triangular system
\begin{equation}
 (I+M)q=v
 \label{eq:chunk-system}
\end{equation}
After solving \eqref{eq:chunk-system}, the states can be reconstructed as
\begin{equation}
 h_t=f_t-\sum_{j=1}^{t}\beta_j E_{t,j}k_jq_j
 \label{eq:chunk-reconstruction}
\end{equation}
This is an exact rearrangement of the recurrence, not an approximation.

The derivation above describes one channel. All $P$ channels in a head share $d_t,\beta_t,k_t$, but have their own input injections and incoming states. Consequently, the coefficient matrix $I+M$ is shared, and the system can be solved jointly for a $c\times P$ matrix of channel-specific right-hand sides.

Rather than explicitly reconstructing the state at every time step, we substitute \eqref{eq:chunk-reconstruction} into the readout to obtain
\begin{equation}
\begin{aligned}
 y_t={}&\widetilde C_t^\top f_t+D u_t\\
 &{}-\sum_{j=1}^t\beta_jE_{t,j}
 (\widetilde C_t^\top k_j)q_j
\end{aligned}
\label{eq:compact-readout}
\end{equation}
The injection terms in $f_t$ are also expanded into products of a vector and a scalar in \eqref{eq:frame}. Computing their inner products with $\widetilde C_t$ and $k_t$ first allows the $n$-dimensional inner products to be shared across $P$ channels. Only the final state of a chunk is reconstructed using \eqref{eq:chunk-reconstruction} and passed to the next chunk.

\section{Task and Training Settings}
\label{app:task_settings}

Repeated training runs use different random seeds, and training batches are generated afresh at each update. Evaluation uses the final checkpoint. For the shell game, direct training is the main comparison. We also evaluate a swap-count curriculum as a training control and report the results in Table~\ref{tab:selected}.

\subsection{Group word problems}
The tasks in Section~\ref{sec:word} use training sequences of up to 64 elements and batch size 64, with between 20,000 and 60,000 updates depending on the model and task. All compared models use direct training at length 64 for $\mathbb{Z}_5$ and a sequence-length curriculum for $S_5$ (generators). In the latter, the length increases from 8 to 64 during the first 60\% of updates and remains at 64 thereafter. Inputs are sampled independently and uniformly from the five increments of $\mathbb{Z}_5$ or the four adjacent transpositions of $S_5$. The training loss includes predictions at every time step. Since every adjacent transposition is odd, the possible labels for $S_5$ (generators) alternate between the alternating group and its coset according to position parity. We therefore use uniform prediction over the 60 candidates allowed by parity as a baseline, with accuracy $1/60$. This does not assume that finite-length prefix products are uniformly distributed over those candidates.

\subsection{Shell-game task}
The fixed-timing frame allocation is given in Section~\ref{sec:shell}. Each observation has 16 components: five cup $x$ coordinates, five $y$ coordinates, a five-dimensional one-hot reveal signal, and one cue. The action has two components. The initial ball location is sampled uniformly from the five slots, and each swap pair is sampled uniformly from the four adjacent pairs. For a fixed-timing episode with $K$ swaps, $K$ of the $W$ windows are selected uniformly without replacement. Each swap follows a smooth eight-step trajectory, with one cup lifted along a sinusoidal arc. The loss is squared error over all time steps, with a weight of 20 on steps after the cue.

All shell-game models use batch size 32 and 20,000 training updates. In curriculum training, the maximum number of swaps increases from 1 to 16 during the first 60\% of updates and remains at 16 thereafter. For the fixed-timing comparison, each trained model is evaluated on 4,096 episodes in each of the standard ($T=213$, $K\le16$) and extended ($T=405$, $K\le32$) conditions. The evaluation swap count is sampled uniformly from zero to the corresponding maximum. For each episode, the horizontal output is averaged over the 12 response frames before selecting the nearest slot. Success is reported by $K$. Overall success is the fraction of successful episodes across the entire evaluation set.

\subsection{Timing-jitter shell game}
Each training episode has 16 windows and a uniformly sampled swap count from zero to 16. Each window's motion duration is sampled independently from $U(4,16)$ and rounded to the nearest integer frame. Waiting durations before each window and before the final cue are independently sampled from $U(0,8)$ with probability 0.8 or $U(16,64)$ otherwise, and rounded in the same way. The reveal and response phases remain five and 12 frames. Padding is excluded from the loss, which is normalized by each episode's valid length before averaging over the batch. The success criterion is the same as in the fixed-timing setting. At evaluation, the window count equals the swap count for 16, 32, 48, 64, 96, and 128 swaps.

\section{Model Configurations}
\label{app:model_settings}

\subsection{Mamba-3 configurations}
Standard Mamba-3 uses the unmodified single-input single-output (SISO) block.\footnote{\url{https://github.com/state-spaces/mamba}} The proposal adds the reflection update to one such block. Both use $d_{\rm model}{=}64$, $d_{\rm state}{=}16$, head dimension 16, and expansion factor 4, with chunk length 64 and four RoPE coordinate pairs. RoPE, the trapezoidal correction, and $B/C$ normalization remain enabled. The two- and four-block configurations change only the number of blocks. The proposed model uses $H=16$ heads with $P=16$ channels per head and $n=16$ state elements per channel, and $a_{\min}=10^{-4}$. Its recurrent state contains 4,096 scalars ($HPn$), excluding auxiliary caches. Mamba-3 + NPLR and standard Mamba-3 use the same task-specific input and output mappings. Group tasks linearly embed one-hot tokens and use a linear classification head. For the shell game, the 16-dimensional observation is linearly mapped to a 64-dimensional input. The model output is linearly mapped to 16 dimensions, and its first two components form the action. These input and output projections have no bias.
For all tasks, these models use Adam with learning rate $10^{-3}$, zero weight decay, cosine decay to zero, and gradient-norm clipping at 1. For all Mamba-3 configurations, $\mathbb{Z}_5$ training uses 20,000 updates and $S_5$ (generators) uses 60,000 updates. The proposed model is trained without early stopping.

\subsection{DeltaProduct configuration}
\label{app:dp_settings}
We use the authors' one-layer block\footnote{\url{https://github.com/automl/DeltaProduct}. We also use the GDN implementation included in this repository.} with model dimension 128, four heads, state dimension 32 per head, and $n_h=2,4$. The decay gate is disabled, and the coefficient range allows negative eigenvalues. We also evaluate GDP~\cite{deltaproduct}, which adds the decay gate to the same configuration (Appendix~\ref{app:dp_gdp}). Group tasks use token embeddings and an output head sized to the number of classes. The shell game uses an input projection from 16-dimensional observations and a two-dimensional output head. Normalization, residual connections, and the multilayer perceptron (MLP) follow the authors' implementation. We use Adam with learning rate $10^{-3}$, zero weight decay, cosine decay to zero, and a gradient-norm limit of 1. Training uses 20,000 updates for $\mathbb{Z}_5$ and the shell game, and 60,000 for $S_5$ (generators), without early stopping.

\subsection{Gated DeltaNet configuration}
We use a publicly available GDN block with one layer, model dimension 128, four heads, key dimension 32 per head, and value dimension 64 per head. The short convolution, output gate, normalization, residual connections, and MLP are retained. Only the coefficient range is extended to $\beta\in(0,2)$. We report this extended variant because it improves accuracy on $S_5$ (generators) over the original range $(0,1)$ under the same training conditions. We use the chunkwise kernel even for short training sequences. Including the task-specific input and output projections, the shell-game model has 333,896 parameters, 8,192 recurrent-state elements, and 2,048 short-convolution cache elements. Group tasks use token embeddings and a linear classification head. Optimization settings and update budgets match those of DeltaProduct. All main comparisons use three runs. Shell-game models use direct training, with the same weights evaluated in the standard and extended conditions.

\subsection{IDS4 configuration}
We use the public code for IDS4~\cite{merrill} with one layer and the default dimensions $d_{\rm model}=255$ and $d_{\rm state}=110$ from its public training configuration.\footnote{\url{https://github.com/jopetty/word-problem}} AdamW uses constant learning rate $10^{-4}$ and weight decay 0.01. Group tasks use token embeddings and a linear classification head, with batch size 64 and 60,000 updates. The shell game uses batch size 32 and 20,000 updates. Each condition is evaluated over three runs.

\subsection{Model sizes and state-update costs}
Table~\ref{tab:sym} summarizes model sizes for the shell game. The proposed model, one-block Mamba-3, DeltaProduct, and GDP have the same recurrent-state size, but their parameter counts differ. State-update costs cover only applying the transition to the previous state, excluding coefficient generation, input injection, and readout. Here $R=1$ is the added rank and $n_h$ is the number of reflection updates per input. For IDS4, the accumulated $110\times110$ matrix gives $S=12{,}100$, and dense matrix multiplication costs $O(S^{3/2})$ per step.

\begin{table}[t]
\centering\footnotesize\setlength{\tabcolsep}{2pt}
\caption{Shell-game model sizes and state-update costs. Parameter counts include task-specific mappings and are rounded to thousands. $S$ counts recurrent-state scalars across all blocks, excluding auxiliary caches. GDN additionally retains 2,048 short-convolution cache elements. Bold identifies the proposed model.}
\label{tab:sym}
\begin{tabular}{lrrl}
\toprule
Model & Params. & State $S$ & \shortstack[l]{State-update\\complexity / step} \\
\midrule
Mamba-3 (1 block) & 57k & 4{,}096 & $O(S)$ \\
Mamba-3 (2 blocks) & 113k & 8{,}192 & $O(S)$ \\
Mamba-3 (4 blocks) & 223k & 16{,}384 & $O(S)$ \\
\textbf{Mamba-3 + NPLR (Proposed)} & \textbf{75k} & \textbf{4{,}096} & $O((R+1)S)$ \\
DeltaProduct ($n_h{=}2$) & 299k & 4{,}096 & $O(n_hS)$ \\
DeltaProduct ($n_h{=}4$) & 366k & 4{,}096 & $O(n_hS)$ \\
GDP ($n_h{=}4$) & 366k & 4{,}096 & $O(n_hS)$ \\
Gated DeltaNet ($\beta\in(0,2)$) & 334k & 8{,}192 & $O(S)$ \\
IDS4 & 3185k & 12{,}100 & $O(S^{3/2})$ \\
\bottomrule
\end{tabular}

\end{table}

\section{Additional Evaluation Results}
\label{app:additional_results}

\subsection{Shell-game training comparison}
Table~\ref{tab:selected} reports three-run means for direct training and the swap-count curriculum, without selecting between them. The proposed model's and IDS4's shell-game entries in Table~\ref{tab:capability} use the standard-condition results after direct training.

Training procedures also affect performance (Table~\ref{tab:selected}). The swap-count curriculum improves mean success rates for Mamba-3 and the GDN extension in the standard condition. However, even four-block Mamba-3 still makes errors in the extended condition. The proposed model achieves perfect success in both conditions under either training procedure. For DeltaProduct, direct training yields higher mean success rates in the extended condition. Under direct training, one IDS4 run predicts only the central slot in both the standard and extended conditions. The other two runs also make errors without swaps ($K=0$). These failures are not limited to composing swap operations and may involve difficulty retaining the initially revealed location or mapping it to the output.

\begin{table}[t]
  \centering\footnotesize\setlength{\tabcolsep}{1.5pt}
  \caption{Overall shell-game success rates (\%, three-run mean) under direct training and the swap-count curriculum (Curr.). Standard uses $K\le16$ and extended uses $K\le32$. All configurations use 20,000 updates. Bold values satisfy $\ge 95\%$ before rounding. $^{\ddagger}$Under direct training, GDN has two perfect runs in the standard condition and two near-perfect runs in the extended condition. The remaining run predicts only the central slot in both conditions.}
  \label{tab:selected}
  \begin{tabular}{lrrrr}
\toprule
& \multicolumn{2}{c}{Standard} & \multicolumn{2}{c}{Extended} \\
\cmidrule(lr){2-3}\cmidrule(lr){4-5}
Model & Direct & Curr. & Direct & Curr. \\
\midrule
Mamba-3 (1 block) & 43.33 & 52.53 & 25.54 & 28.35 \\ %
Mamba-3 (2 blocks) & 57.55 & 83.97 & 37.45 & 55.12 \\ %
Mamba-3 (4 blocks) & 74.65 & 90.50 & 49.31 & 63.40 \\ %
\textbf{Mamba-3 + NPLR (Proposed)} & \textbf{100.00} & \textbf{100.00} & \textbf{100.00} & \textbf{100.00} \\ %
DeltaProduct ($n_h{=}2$) & \textbf{99.99} & \textbf{99.88} & \textbf{99.51} & \textbf{96.46} \\ %
DeltaProduct ($n_h{=}4$) & \textbf{99.99} & \textbf{99.69} & \textbf{99.36} & 91.83 \\ %
Gated DeltaNet ($\beta\in(0,2)$) & 73.62$^{\ddagger}$ & \textbf{99.99} & 73.23$^{\ddagger}$ & 84.95 \\ %
IDS4 & 28.03 & 29.00 & 21.73 & 20.64 \\ %
\bottomrule
\end{tabular}

\end{table}

\subsection{Depth comparison}
For the two- and four-block Mamba-3 configurations in Table~\ref{tab:nplr}, final-position accuracies on $S_5$ (generators) at $L=64$ range from 3.52\% to 6.64\% and from 47.46\% to 61.91\%, respectively, across three runs.

\subsection{Length extrapolation}
To assess variation in length extrapolation, all five configurations in Fig.~\ref{fig:extrap} use five runs on $S_5$ (generators), whereas $\mathbb{Z}_5$ uses three. For each task in Table~\ref{tab:extrapdetail}, all runs use the same 512 sequences at each evaluation length, generated independently of training.

\begin{table}[t]
\centering\footnotesize
\caption{Length extrapolation of Mamba-3 + NPLR and its no-RoPE ablation. Values are final-position accuracy (\%). All models are trained at length 64. Mean is the arithmetic mean. Min and Max are the extremes across runs. Bold values satisfy $\ge 95\%$ before rounding.}
\label{tab:extrapdetail}
\begin{tabular}{rrrr}
\toprule
Length $L$ & Mean & Min & Max \\
\midrule
\multicolumn{4}{l}{(a) $\mathbb{Z}_5$, proposed (3 runs)} \\
64 & \textbf{100.00} & \textbf{100.00} & \textbf{100.00} \\
128 & \textbf{100.00} & \textbf{100.00} & \textbf{100.00} \\
256 & 94.40 & 87.30 & \textbf{100.00} \\
512 & 66.15 & 49.02 & 78.52 \\
1024 & 42.06 & 20.51 & 72.46 \\
\midrule
\multicolumn{4}{l}{(b) $S_5$, proposed (5 runs)} \\
64 & \textbf{100.00} & \textbf{100.00} & \textbf{100.00} \\
128 & \textbf{99.77} & \textbf{98.83} & \textbf{100.00} \\
256 & 82.66 & 24.41 & \textbf{100.00} \\
512 & 55.31 & 11.33 & \textbf{100.00} \\
1024 & 19.61 & 3.12 & 53.12 \\
2048 & 3.16 & 1.95 & 6.45 \\
\midrule
\multicolumn{4}{l}{(c) $S_5$, w/o RoPE (5 runs)} \\
64 & \textbf{100.00} & \textbf{100.00} & \textbf{100.00} \\
128 & \textbf{100.00} & \textbf{100.00} & \textbf{100.00} \\
256 & \textbf{100.00} & \textbf{100.00} & \textbf{100.00} \\
512 & \textbf{100.00} & \textbf{100.00} & \textbf{100.00} \\
1024 & 91.52 & 57.62 & \textbf{100.00} \\
2048 & 67.30 & 4.49 & \textbf{100.00} \\
\bottomrule
\end{tabular}

\end{table}

\subsection{Success rates by swap count}
Tables~\ref{tab:shell_full_standard} and~\ref{tab:shell_full_extended} give the three-run means at intervals of four swaps from Fig.~\ref{fig:rx}, together with GDP, GDN, and IDS4 results. All models use 20,000 direct-training updates and the same 4,096 evaluation episodes per run and condition. Overall includes episodes at every integer swap count, including counts omitted from the tables, and is averaged over episodes within each run. The no-tracking prior is analytical.

\begin{table*}[t]
\centering\footnotesize\setlength{\tabcolsep}{3pt}
\caption{Standard shell-game success rates (\%, three-run mean) at intervals of four swaps ($T=213$, $K\le16$). Overall covers the full range of $K$. Bold values satisfy $\ge95\%$ before rounding. $^{\dagger}$One failure at $K=15$ across the three runs gives an overall success rate of 99.99\%, despite perfect success at the displayed swap counts. $^{\ddagger}$GDN and GDP each have two runs with 100\% overall success and one run that predicts only the central slot.}
\label{tab:shell_full_standard}
\begin{tabular}{lrrrrrr}
\toprule
Model & $K=0$ & $K=4$ & $K=8$ & $K=12$ & $K=16$ & Overall \\
\midrule
No-tracking prior & \textbf{100.00} & 38.28 & 30.98 & 27.11 & 24.72 & --- \\
\midrule
Mamba-3 (1 block) & \textbf{100.00} & 48.76 & 37.18 & 29.88 & 23.83 & 43.33 \\
Mamba-3 (2 blocks) & \textbf{100.00} & 78.95 & 50.61 & 34.22 & 31.69 & 57.55 \\
Mamba-3 (4 blocks) & \textbf{100.00} & 94.95 & 75.52 & 54.25 & 45.98 & 74.65 \\
\textbf{Mamba-3 + NPLR (Proposed)} & \textbf{100.00} & \textbf{100.00} & \textbf{100.00} & \textbf{100.00} & \textbf{100.00} & \textbf{100.00} \\
DeltaProduct ($n_h{=}2$) & \textbf{100.00} & \textbf{100.00} & \textbf{100.00} & \textbf{100.00} & \textbf{99.87} & \textbf{99.99} \\
DeltaProduct ($n_h{=}4$) & \textbf{100.00} & \textbf{100.00} & \textbf{100.00} & \textbf{100.00} & \textbf{100.00} & \textbf{99.99}$^{\dagger}$ \\
GDP ($n_h{=}4$)$^{\ddagger}$ & 73.47 & 74.06 & 74.70 & 74.93 & 73.52 & 73.62 \\
Gated DeltaNet ($\beta\in(0,2)$)$^{\ddagger}$ & 73.47 & 74.06 & 74.70 & 74.93 & 73.52 & 73.62 \\
IDS4 & 60.66 & 29.19 & 25.45 & 27.59 & 23.58 & 28.03 \\
\bottomrule
\end{tabular}

\end{table*}

\begin{table*}[t]
\centering\footnotesize\setlength{\tabcolsep}{3pt}
\caption{Extended shell-game success rates (\%, three-run mean) at intervals of four swaps ($T=405$, $K\le32$). Training uses $K\le16$. Overall covers the full range of $K$. Bold values satisfy $\ge95\%$ before rounding. $^{\ddagger}$GDN and GDP each have two near-perfect runs and one run that predicts only the central slot.}
\label{tab:shell_full_extended}
\begin{tabular}{lrrrrrrrrrr}
\toprule
Model & $K=0$ & $K=4$ & $K=8$ & $K=12$ & $K=16$ & $K=20$ & $K=24$ & $K=28$ & $K=32$ & Overall \\
\midrule
No-tracking prior & \textbf{100.00} & 38.28 & 30.98 & 27.11 & 24.72 & 23.15 & 22.11 & 21.41 & 20.94 & --- \\
\midrule
Mamba-3 (1 block) & 70.08 & 34.60 & 26.79 & 21.82 & 19.21 & 20.72 & 19.39 & 24.18 & 22.01 & 25.54 \\
Mamba-3 (2 blocks) & 94.35 & 66.68 & 41.63 & 31.57 & 24.39 & 25.75 & 21.40 & 23.05 & 22.95 & 37.45 \\
Mamba-3 (4 blocks) & \textbf{100.00} & 88.17 & 69.61 & 46.82 & 35.21 & 33.61 & 24.83 & 22.51 & 22.23 & 49.31 \\
\textbf{Mamba-3 + NPLR (Proposed)} & \textbf{100.00} & \textbf{100.00} & \textbf{100.00} & \textbf{100.00} & \textbf{100.00} & \textbf{100.00} & \textbf{100.00} & \textbf{100.00} & \textbf{100.00} & \textbf{100.00} \\
DeltaProduct ($n_h{=}2$) & \textbf{100.00} & \textbf{100.00} & \textbf{100.00} & \textbf{100.00} & \textbf{100.00} & \textbf{100.00} & \textbf{99.73} & \textbf{97.80} & \textbf{97.07} & \textbf{99.51} \\
DeltaProduct ($n_h{=}4$) & \textbf{100.00} & \textbf{100.00} & \textbf{100.00} & \textbf{100.00} & \textbf{100.00} & \textbf{99.72} & \textbf{99.44} & \textbf{97.43} & \textbf{96.49} & \textbf{99.36} \\
GDP ($n_h{=}4$)$^{\ddagger}$ & 72.88 & 72.36 & 75.47 & 73.33 & 74.31 & 73.70 & 73.53 & 73.11 & 73.02 & 73.35 \\
Gated DeltaNet ($\beta\in(0,2)$)$^{\ddagger}$ & 72.88 & 72.36 & 75.02 & 73.05 & 74.31 & 73.70 & 73.53 & 73.11 & 73.28 & 73.23 \\
IDS4 & 18.79 & 24.40 & 25.15 & 22.58 & 20.21 & 17.73 & 20.07 & 21.08 & 22.82 & 21.73 \\
\bottomrule
\end{tabular}

\end{table*}

\subsection{Supplemental comparison with Gated DeltaProduct}
\label{app:dp_gdp}
To test whether the timing-jitter advantage persists when the decay gate is enabled, we evaluate the GDP configuration specified in Appendix~\ref{app:dp_settings}. Block dimensions and evaluation protocols otherwise match DeltaProduct. Training settings are reused from DeltaProduct without GDP-specific tuning. Figure~\ref{fig:shell_jitter} shows the timing-jitter results, and Table~\ref{tab:dp_gdp} summarizes the group tasks and fixed-timing shell game. All three runs are included.

Under direct training, adding the decay gate led to greater variation across runs in the fixed-timing shell game with $n_h=4$, including one run that converged to a constant prediction. This suggests reduced training stability under the evaluated conditions, although the curriculum improved performance (Table~\ref{tab:dp_gdp}). One timing-jitter run also predicted only the central slot. All runs are included in the reported means. The proposed model's advantage on longer timing-jitter sequences persists when the decay gate is enabled (Fig.~\ref{fig:shell_jitter}).

\begin{table*}[t]
\centering\footnotesize\setlength{\tabcolsep}{4pt}
\caption{DeltaProduct and GDP under the same training and evaluation protocols (\%, three-run means). Group-task columns give final-position accuracy. Shell-game columns give overall success under direct training or the swap-count curriculum (Curr.). All runs are included. Bold values satisfy $\ge95\%$ before rounding.}
\label{tab:dp_gdp}
\begin{tabular}{lrrrrrrr}
\toprule
& $\mathbb Z_5$ & \multicolumn{2}{c}{$S_5$ (generators)} & \multicolumn{2}{c}{Standard shell game} & \multicolumn{2}{c}{Extended shell game} \\
Model & \multicolumn{1}{c}{$L=64$} & \multicolumn{1}{c}{$L=64$} & \multicolumn{1}{c}{$L=128$} & \multicolumn{1}{c}{Direct} & \multicolumn{1}{c}{Curr.} & \multicolumn{1}{c}{Direct} & \multicolumn{1}{c}{Curr.} \\
\midrule
DeltaProduct ($n_h=2$) & 89.78 & \textbf{100.00} & \textbf{99.28} & \textbf{99.99} & \textbf{99.88} & \textbf{99.51} & \textbf{96.46} \\
DeltaProduct ($n_h=4$) & \textbf{99.48} & \textbf{100.00} & \textbf{99.87} & \textbf{99.99} & \textbf{99.69} & \textbf{99.36} & 91.83 \\
GDP ($n_h=2$) & 94.60 & \textbf{100.00} & \textbf{99.54} & \textbf{99.99} & \textbf{99.82} & \textbf{99.50} & 94.67 \\
GDP ($n_h=4$) & \textbf{97.46} & \textbf{100.00} & \textbf{100.00} & 73.62 & \textbf{100.00} & 73.35 & \textbf{97.93} \\
\bottomrule
\end{tabular}
\end{table*}

\FloatBarrier

\end{document}